\documentclass[sigconf]{acmart}

\def\BibTeX{{\rm B\kern-.05em{\sc i\kern-.025em b}\kern-.08emT\kern-.1667em\lower.7ex\hbox{E}\kern-.125emX}}
    
\copyrightyear{2019}
\acmYear{2019}
\setcopyright{acmlicensed}
\acmConference[The Web Conference '19]{The Web Conference 2019}{May 13--17, 2019}{San Francisco, CA}
\acmBooktitle{The Web Conference 2019, May 13--17, 2019, San Francisco, CA}
\acmPrice{15.00}
\acmDOI{}
\acmISBN{}

\begin{document}

%
\title{Bot Electioneering Volume: Visualizing~Social~Bot~Activity~During~Elections}

%
\author{Kai-Cheng Yang}
\email{yangkc@iu.edu}
\orcid{0000-0003-4627-9273}
\affiliation{
  \institution{Center for Complex Networks and Systems Research, Indiana University}
  \city{Bloomington}
  \state{Indiana}
  \postcode{47408}
}

\author{Pik-Mai Hui}
\email{huip@iu.edu}
\affiliation{
  \institution{Center for Complex Networks and Systems Research, Indiana University}
  \city{Bloomington}
  \state{Indiana}
  \postcode{47408}
}

\author{Filippo Menczer}
\email{cnets.indiana.edu/fil}
\orcid{0000-0003-4384-2876}
\affiliation{
  \institution{Center for Complex Networks and Systems Research, Indiana University}
  \city{Bloomington}
  \state{Indiana}
  \postcode{47408}
}

%
\renewcommand{\shortauthors}{Yang, et al.}

%
\begin{abstract}
It has been widely recognized that automated bots may have a significant impact on the outcomes of national events. It is important to raise public awareness about the threat of bots on social media during these important events, such as the 2018 US midterm election. To this end, we deployed a web application to help the public explore the activities of likely bots on Twitter on a daily basis. The application, called Bot Electioneering Volume (BEV), reports on the level of likely bot activities and visualizes the topics targeted by them. With this paper we release our code base for the BEV framework, with the goal of facilitating future efforts to combat malicious bots on social media.
\end{abstract}

%
\maketitle

\section{Introduction}

Social bots have drawn great attention from the public recently.
They are accounts on social media platforms controlled at least in part by algorithms to generate/share/retweet content and interact with human users \cite{ferrara2016rise}.
The automated nature of social bots makes it easy to achieve scalability, with which a single person is capable to control  thousands of accounts on one or more social media platforms.
When needed, these social bots can work collectively to manipulate the public by promoting certain accounts or opinions.

Being social animals, human users are inevitably vulnerable to the efforts of the social bots.
Studies have shown ubiquitous social bots \cite{varol2017online} distort online discussions, and particularly those about politics.
During the 2010 US midterm election, primitive social bots were used to attack some candidates \cite{metaxas2012social} and spread tweets with links to fake news websites \cite{ratkiewicz2011detecting}.
A similar pattern emerged in the 2016 US presidential election, only with more sophisticated bots that aimed to effectively push their messages to the target audience \cite{bessi2016social}. In particular, bots were most active in the core of the misinformation-sharing network \cite{Shao2018anatomy} and effectively amplified the spread of low-credibility content by posting it within seconds and by targeting influential accounts \cite{Shao18hoaxybots}. 
Analogous automated campaigns were reported in countries around the globe \cite{stella2018bots,ferrara2017disinformation}. 

\begin{figure}[b]
    \centering
    \includegraphics[width=0.45\textwidth]{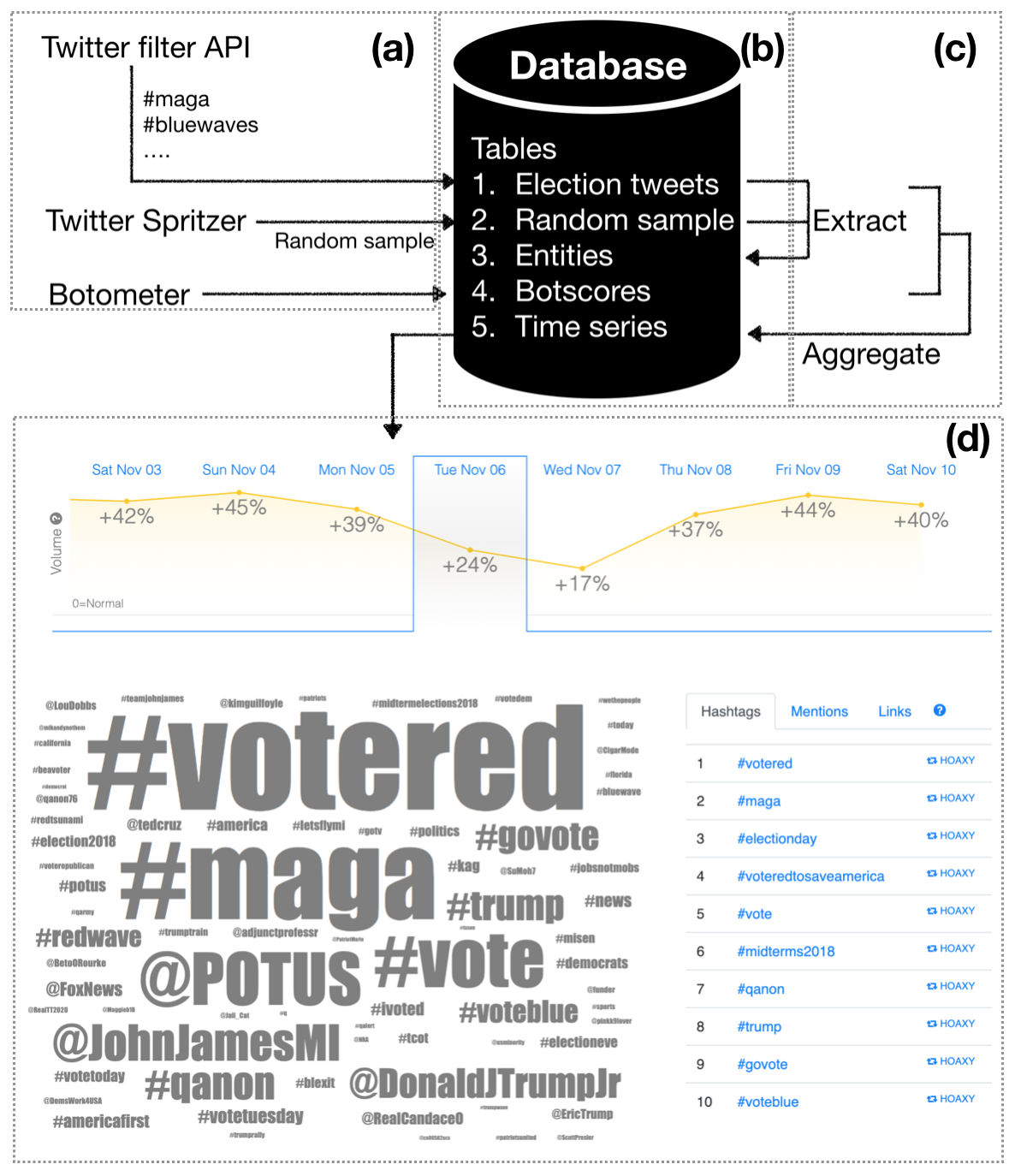}
    \caption{Illustration of BEV's (a)~crawler, (b)~database, (c)~analyzer, and screen shot of (d)~front-end interface. The upper panel of the frontend shows the Bot Electioneering Volume for the past 8 days. Users can select a day of interest to explore the top topics of that day. The bottom panels show a tag cloud and entity lists for the selected day. The tag cloud presents entities all together, with the size of each entity proportional to how often it is tweeted by bots. The entity lists display hashtags, mentions, and links ranked by how often each is tweeted by bots.}
    \label{fig:system_design_screen_shot}
\end{figure}

Here we present Bot Electioneering Volume (BEV), a platform that visualizes the volume generated by bots and the corresponding targeted topics. 
BEV tracks online traffic centered around elections from Twitter by feeding the streaming API with a list of selected hashtags.
By incorporating the bot-detection ability of Botometer \cite{davis2016botornot,varol2017online,yang2019arming}, BEV is able to distinguish between content generated by likely bots and humans.
The measurement of average bot activity is then compared with random samples of tweets to produce a number that quantifies electioneering activity by bots.
BEV also collects content topics, including hashtags, mentions, and URLs shared by likely bots, and reports on their relative volumes.


BEV (\href{https://botometer.iuni.iu.edu/bev/}{botometer.iuni.iu.edu/bev}) monitored public tweets about the 2018 US midterm elections between October 22, 2018 and Dec 30, 2018.
During the collection period, BEV drew over 3,000 visits.
An archive of the data from Oct to Dec 2018 remains publicly available for retrospective inspections. 
We plan to activate BEV again for future elections.
Our goal is to raise public awareness of bot activities and their impact during elections in the past, and more importantly those in the future.

\section{System design}

The BEV system 
contains 4 major parts: a crawler, a database, an analyzer, and a front-end interface, as illustrated in Figure~\ref{fig:system_design_screen_shot}.

The crawler is in charge of tracking Twitter's filtering API for public election-related tweets, querying Twitter's Spritzer API for random samples of public tweets, and fetching bot scores. 
Crawled data is stored in the database. 
The analyzer then extracts the required information and generates the statistics for the visualization at the application frontend.
The frontend has three major parts: the Bot Electioneering Volume timeline, a tag cloud, and entity lists.
The Bot Electioneering Volume measures the activity of likely bots, while the tag could and entity lists display the topics that are most tweeted by likely bots.
By clicking on the links in an entity list, users are directed to Hoaxy (\href{https://hoaxy.iuni.iu.edu/}{hoaxy.iuni.iu.edu}) \cite{shao2016hoaxy}, where they can explore more in-depth visualizations of the influence of bots around the entities on Twitter in the recent minutes/hours/days. 

The data collection runs in a streaming fashion, but fetching bot scores and analyzing the data take time.
The front-end interface is updated every 4 hours to reflect newly incoming data.

\subsection{Data collection}

The collection of election-related tweets is crucial to our application.
Our collection process starts with a set of election-related hashtags that are tracked using the Twitter filtering API.
We seeded our set of hashtags with several widely-used political hashtags including \#2018midterms, \#maga, and \#bluewave.
We then repeatedly expanded the set with co-occuring hashtags \cite{conover12partisan}, resulting in a set of 110 hashtags.
From this set we manually removed 6 hashtags that are general and irrelevant to election.
We also added the hashtags for each US state's Senate race: \#casen, \#nysen, and so on.
The full list of hashtags can be found in the FAQ page of the BEV website. 
This methodology allows our system to collect most tweets with newly emerging election hashtags, because it is likely that these tweets contain some of the hashtags in our list as well.

Twitter's free filtering API offers at most 1\% of Twitter's traffic.
We estimated that the traffic captured by our method is about 0.3--0.5\% of Twitter's complete traffic (see Figure~\ref{fig:bev}(a)).
This means that instead of a sample, BEV collects and visualizes bot activities based on \emph{all} of the targeted election-related tweets. 

\subsection{Bot identification}

BEV uses Botometer (\href{https://botometer.iuni.iu.edu/}{botometer.iuni.iu.edu}) \cite{davis2016botornot,varol2017online,yang2019arming} to obtain bot scores for Twitter accounts involved in election-related discourse.
Botometer is a supervised machine learning algorithm that considers more than a thousand features about an account and its activity to estimate the likelihood that the account is automated.
We consider accounts with bot score above 4 (on a 5-point scale) as social bots. This is a fairly conservative threshold choice, corresponding to a posterior probability of automation near 50\% \cite{yang2019arming} based on a 15\% prior probability \cite{varol2017online}.

\subsection{Bot activity measurement}

\begin{figure}
    \centering
    \includegraphics[width=0.45\textwidth]{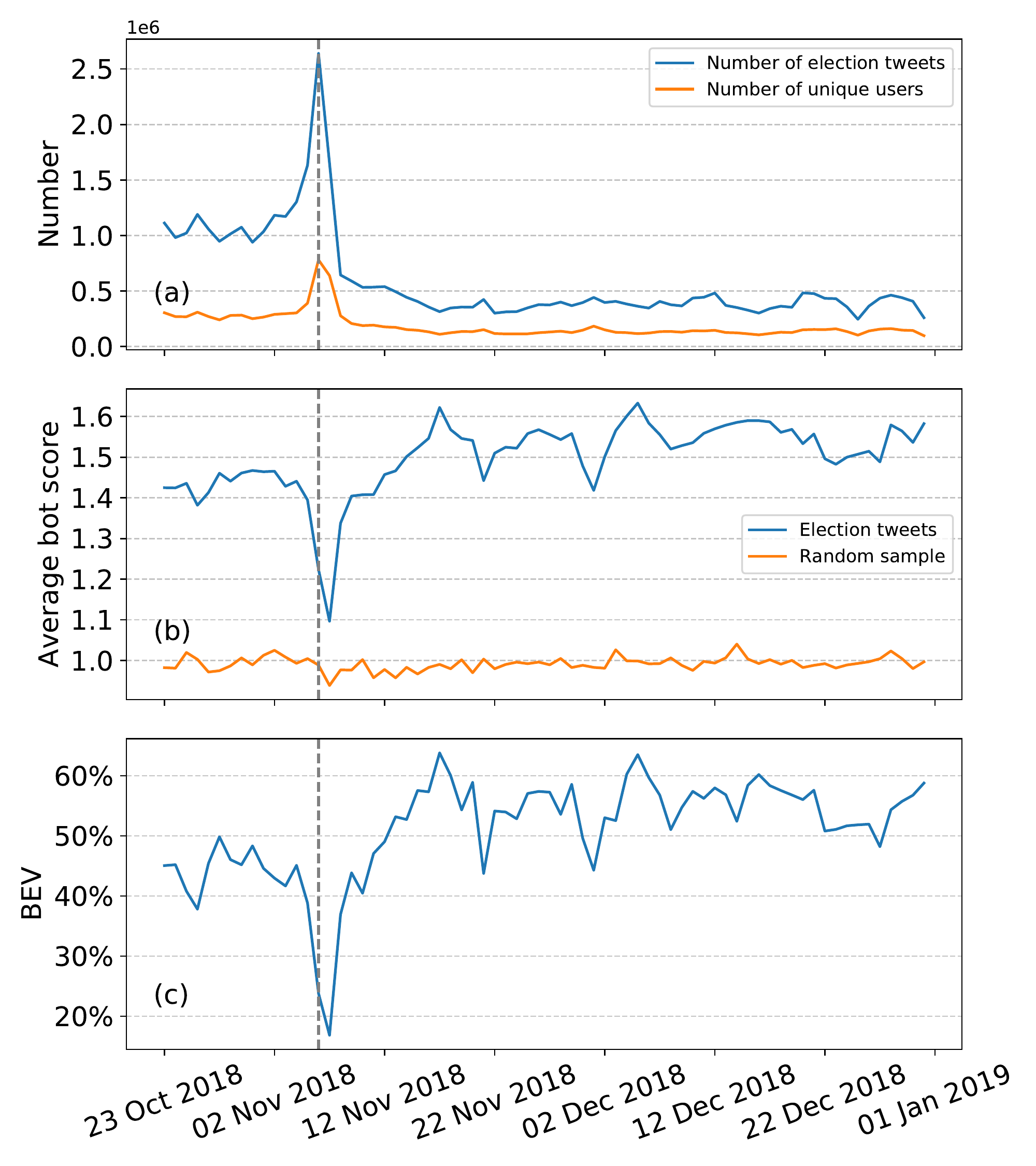}
    \caption{(a)~Number of election tweets and unique users for each day. (b)~Average bot scores for election tweets and random sample. (c)~BEV timeline. 
    The 2018 US midterm elections day, Nov 6, is highlighted by a vertical dashed line.}
    \label{fig:bev}
\end{figure}

To measure the bot electioneering volume, we first take daily averages of the bot scores of accounts generating political tweets, weighted by their tweet frequencies.
Considering that spamming the same message is a common strategy for bots, the weighted average better highlights the amount of bot-generated content and their potential influence.
To obrain a baseline of bot activity, we produce the same weighted average for random tweets from Twitter's Spritzer API, with a rate of 1,000 tweets per hour.
The daily average bot scores for election and random tweets are shown in Figure~\ref{fig:bev}(b).

The Bot Electioneering Volume is defined by the relative difference between the two averages:
\begin{equation}
    \text{BEV} = \frac{S_\text{Electoral} - S_\text{RandomSample}}{S_\text{RandomSample}} \label{eq:bev_formula}\text{,}
\end{equation}
where $S_\text{Electoral}$ and $S_\text{RandomSample}$ represent the average bot scores of electoral and random tweets, respectively.
BEV is shown as a percentage difference in the front-end interface, as shown in Figure~\ref{fig:system_design_screen_shot}(d). Figure~\ref{fig:bev}(c) plots the Bot Electioneering Volume of the whole midterm elections period.

\begin{figure}
    \centering
    \includegraphics[width=0.45\textwidth]{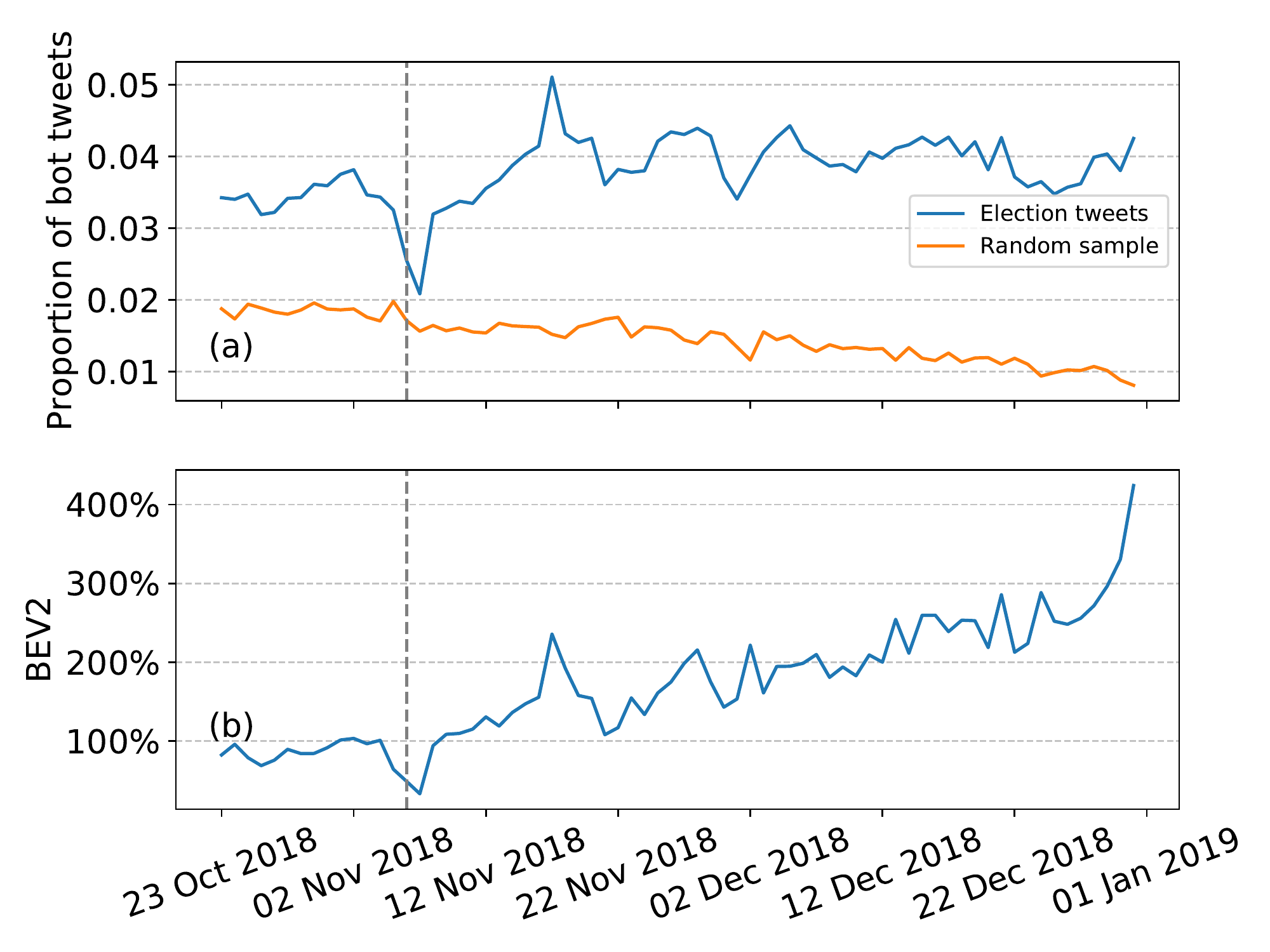}
    \caption{(a) Proportion of tweets by bots for election tweets and random sample. (b) BEV2.}
    \label{fig:bev2}
\end{figure}

In the design stage of BEV, we considered different metrics by replacing the average bot score $S$ in Eq.~\ref{eq:bev_formula} with the median score and the proportion of tweets by bots.
The Bot Electioneering Volume based on median values of the scores yields patterns similar to the average score version.
However, the version based on the proportion of tweets by bots (denoted by BEV2) shows different trends.
As shown in Figure~\ref{fig:bev2}(a), the baseline proportion of tweets by bots decreases steadily after the midterm elections. Consequently, the BEV2 value increases (Figure~\ref{fig:bev2}(b)).
Since the proportion of tweets by likely bots in the random sample is not stable, 
we decided to deploy the BEV metric based on average bot scores.

The BEV front-end interface also visualizes the topics targeted by likely bots.
For simplicity, we only extract hashtags, mentioned accounts, and links from the tweets generated by bots.
The targeted topics are represented in a tag cloud and entity lists.

\section{Discussion}

BEV reveals many interesting patterns.
In terms of the discussion intensity, the number of election tweets before the election was about twice as many as the number after the election day peak, as shown in Figure~\ref{fig:bev}(a). 
For bot activity, all the different metrics --- average bot score, median bot score, and proportion of tweets by likely bots --- are much higher for election tweets, suggesting that bots are actively generating election-related content; they are indeed employed for electioneering.

The average bot scores of election-related tweets fluctuated from day to day.
The BEV on November 6 and 7 was drastically lower than on other days.
Considering the spike in election tweets, we hypothesize that the bot activity was diluted by the huge amount of normal users around election day.
Furthermore, the average bot scores of election-related tweets after the election are generally higher than those before the election.
This is perhaps because many human users ceased to tweet political content after the election, but the bots kept working.
BEV cannot reflect the changes in total volume of tweets because it is based on averaging the daily scores.
For the future, a measurement that can leverage both the average bot score and the volume of tweets may be preferable to better represent the volume generated by bots.

\section{Conclusion}

We offer a real-time tool to visualize the electioneering activities of social bots. Open-source code for BEV is at \href{https://github.com/IUNetSci/BEV}{github.com/IUNetSci/BEV}.

Around the 2018 US midterm elections, bot electioneering was rampant. The great majority of content amplified by likely bots was on the conservative side of the political spectrum. It remains to be seen if this will change in the future.

The tool enables the public to gain a sense of the organic nature of the online discussion regarding elections and spot possibly polluted content. 
By performing the hashtag list generation procedure with different seeds, BEV can be adapted to target future US elections, as well as elections and events in other countries.

\paragraph{Acknowledgments.} We are grateful to the IU Network Science Institute (\href{http://iuni.iu.edu/}{iuni.iu.edu}) and the AWS Cloud Credits for Research program for supporting the BEV infrastructure and to Twitter for their free APIs. Botometer was developed with support from the Democracy Fund.

\bibliographystyle{ACM-Reference-Format}
\bibliography{ref}

\end{document}